# A monolithic fabrication platform for intrinsically stretchable polymer transistors and complementary circuits

Yujia Yuan[1†], Chuanzhen Zhao[2†], Margherita Ronchini[1,4†], Yuya Nishio[1], Donglai Zhong[1], Can Wu[1], Hyukmin Kweon[1], Zehao Sun[2], Rachael K. Mow[2], Yuran Shi[2], Lukas Michalek[2], Haotian Wu[3], Qianhe Liu[2], Weichen Wang[2], Yating Yao[5], Zelong Yin[6], Junyi Zhao[2], Zihan He[2], Ke Chen[2], Ruiheng Wu[2], Jiuyun Shi[2], Jian Pei[3] and Zhenan Bao[2*]

[1]Department of Electrical Engineering, Stanford University, Stanford, CA, USA

[2]Department of Chemical Engineering, Stanford University, Stanford, CA, USA

[3]College of Chemistry and Molecular Engineering, Peking University, Beijing, China.

[4]Department of Electrical and Computer Engineering, Aarhus University, Aarhus, Denmark

[5]Department of Chemistry, Stanford University, Stanford, CA, USA

[6]Department of Applied Physics, Stanford University, Stanford, CA, USA

[†]These authors contributed equally to this work
*Corresponding author: zbao@stanford.edu

## ABSTRACT

Soft and stretchable organic field-effect transistors (OFETs) and circuits offer powerful signal-conditioning capabilities for bioelectronic devices with tunable mechanical and chemical properties. However, current fabrication methods are material-dependent, i.e., each newly developed polymer semiconductor (PSC) requires tailored fabrication procedures. This severely limits the rapid development of functional circuits. The difficulty increases substantially further for complementary OFET circuits, as two types of polymer semiconductors must be sequentially patterned, often leading to device degradation. Here, we introduce a universal monolithic photolithography process for producing stretchable complementary OFETs and circuits with high yield and high resolution. This was enabled by a rational process-design conceptual framework involving: a direct, photo-patternable, solvent-resistant, crosslinked dielectric/semiconductor interface; a broadly applicable method for crosslinked polymer semiconductor blends that maintain high mobility; and a new method for polymer semiconductor patterning that involves simultaneous etch masking and encapsulation. We achieve the highest integration density to date for stretchable OTFTs, at 55,000 per $cm^2$, with channel-length resolution down to 2 μm and 5 V low-voltage operation. By engineering two critical interfaces above and underneath the PSC, we provide a versatile, photo-patterning, and encapsulation platform for stretchable OFETs. We demonstrate its usage to pattern various PSC types, including complementary OFET circuits. We demonstrate 3 kHz stretchable complementary OFET ring oscillators, the first to exceed 1 kHz and representing a >60-fold increase in stage switching speed over the current state of the art. Building on this platform, we demonstrate the first stretchable complementary OTFT neuron circuit, in which the output frequency can be modulated by the input current, mimicking neuronal signal processing. The described process, with high-density integration and high scalability, can be readily adapted to incorporate various high-performance stretchable materials. Therefore, it provides a foundational process that may accelerate the development and future manufacturing of skin-like electronics.

**Introduction**

Stretchable electronics such as wearable health monitors[1-6], tissue-like implants[7-12], artificial electronic skin (e-skin)[13-17], and soft intra-operative tools[18-20] have spurred rapid advancements in bioelectronics by bridging the gap between conventional rigid electronics and soft, dynamic biological systems. This technology is a promising candidate for realizing brain-machine interfaces[21-24], augmented reality[25-28], and soft robotics[29-33]. To achieve low stiffness and high stretchability, several strategies have been proposed, including material-level design and engineering[34-38] and device-level strain engineering[39-45]. Among these, intrinsically stretchable polymers have gained significant attention for their robustness due to uniform strain distribution and their potential for high spatial resolution[34,37,46,47].

Intrinsically stretchable polymer materials provide high molecular tunability, solution processability, and biocompatibility, making them ideal for on-skin signal transduction, processing, and communication[15,48-50]. Over the past few years, significant advances have been made in the development of intrinsically stretchable polymer semiconductors (PSCs) and the fabrication processes of OFET arrays and circuits. For example, stretchable transistors with high mobility (~1 $cm^2 v^{-1} s^{-1}$) and stretchability (~100% strain) can be achieved using nanoconfined PSC networks embedded in an elastic matrix[51-53]. To achieve superior mechanical performance, all components in OFETs, including the channel, gate, source, and drain electrodes, dielectrics, and encapsulation, should be intrinsically stretchable. Our group previously reported stretchable OFET fabrication processes and successfully fabricated inverters, ring oscillators (ROs), amplifiers, logic gates, and analog-to-frequency conversion circuits[48,54]. However, these processes required manual alignment of shadow masks for drain and source electrodes, limiting transistor channel lengths to ~70 μm with ~350 OFETs per $cm^2$. Another approach through direct photo-patterning of all active layers was subsequently developed[34], enabling a much higher device density at 42,000 units $cm^{-2}$ at modest mobility (0.25 $cm^2V^{-1}s^{-1}$). However, the operating voltage was relatively high (~30 V).

A significant bottleneck in the field stems from the highly material-dependent nature of existing fabrication techniques. The previously reported processes were designed for specific materials, in which device structures, dielectric interfaces, and patterning strategies depend on the PSCs. The reliance on customized processing protocols for individual PSCs hinders scalability and broader advancement, as the introduction of new high-mobility channel materials invariably requires the development of new lithography and fabrication processes. Additionally, most reported intrinsically stretchable OFETs are constructed with unipolar p-type transistors[34,35,46,48,54]. Complementary circuits, which comprise both p-type and n-type transistors, are advantageous due to their lower static power consumption, increased functionality, and smaller device footprint. There are limited methods reported for stretchable complementary circuit fabrication, which all rely on shadow mask patterning, limiting the channel lengths (~100 μm), throughput, and capability for large-area patterning[55,56]. Therefore, developing a versatile, channel material-independent fabrication platform for high-performance, high-resolution stretchable polymer transistors and complementary circuits is crucial for next-generation skin electronics.

In this work, we report a versatile photolithography-based fabrication platform for large-area, high-resolution stretchable polymer transistors and complementary circuits (**Fig. 1a, 1b**). This work provides a conceptual framework for rational process design by engineering universal dielectric interfaces for both bottom and top dielectric interfaces with PSCs, applicable to a wide range of stretchable polymer semiconductors, including both n-type and p-type semiconductors (**Fig. 1c**). This framework consists of three core aspects: 1) a direct photo-patternable solvent-resistant crosslinked dielectric/semiconductor interface, 2) a broadly applicable method for crosslinked polymer semiconductor blend maintaining high mobility, and 3) a new method for polymer semiconductor patterning involving simultaneous etch masking and encapsulation. As a result, our process can be readily adapted to incorporate various high-performance stretchable materials. We fabricated devices comprising three stretchable polymer-semiconductor blends, each incorporating different types of semiconductors and crosslinkers. Building on this interface engineering, we developed a monolithic sequential fabrication process with a self-aligned encapsulation (SAE) technique that eliminates layer transfers, releases, or manual alignment, resulting in high yield and good repeatability. This process achieves a record-high OFET device density of 55,000 OFETs per cm$^2$ with a channel length of ~2 μm and a low operating voltage of 5 V at 15 nA/μm (current output/channel width) (**Fig. 1b, 1e, Supplementary Table 1**). Stretchable complementary circuits were fabricated, including inverters, amplifiers, and ring oscillators. The stretchable OFET arrays conform to three-dimensional braille patterns on a fingertip (**Fig. 1f**) and exhibit multi-axial stretchability (**Fig. 1g**). Our cleanroom-compatible process is high-throughput and scalable, enabling the fabrication of wafer-scale OFET arrays. This process provides a foundation for accelerating the development and future manufacturing of skin-like electronics.

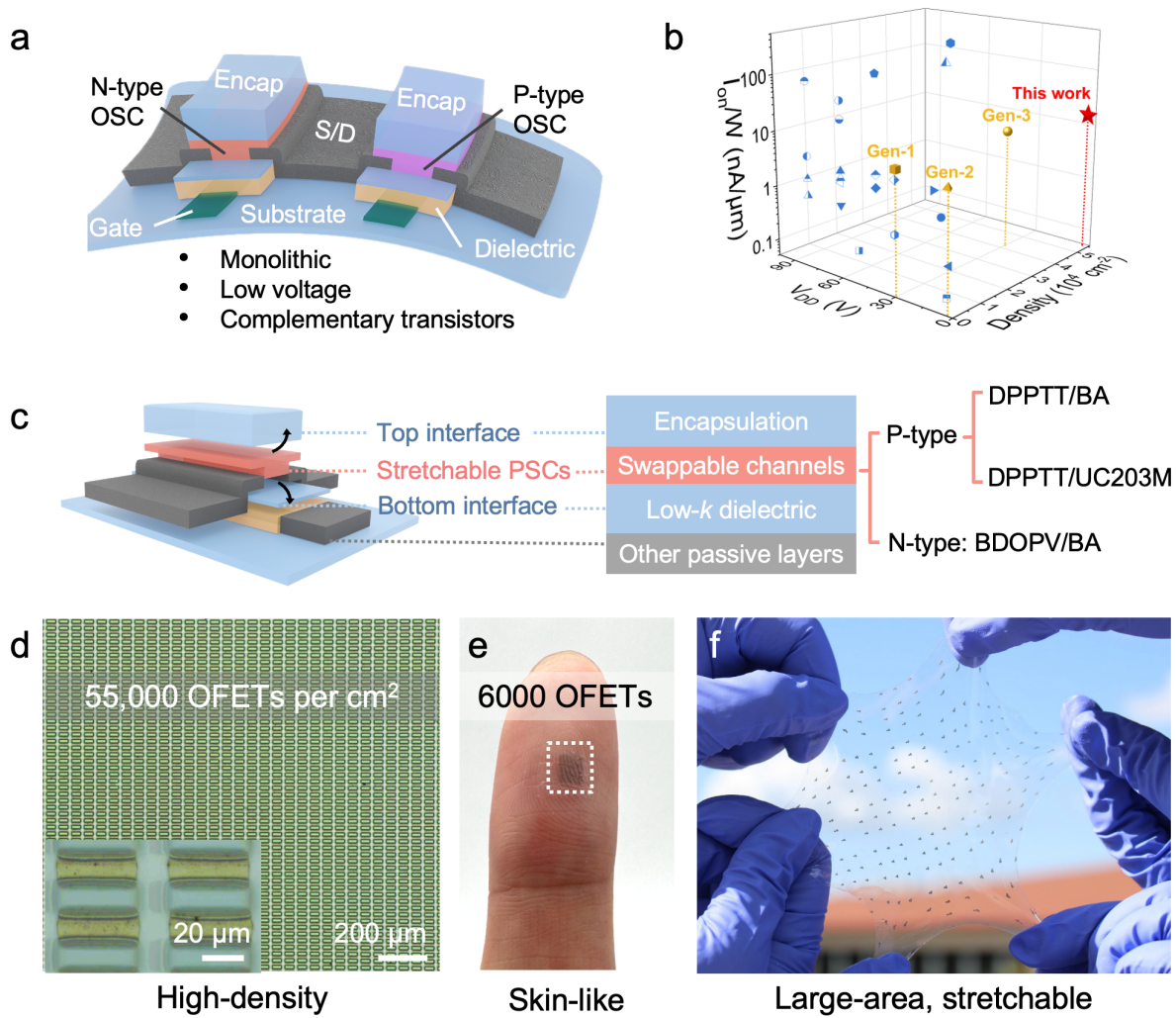

**Fig. 1. Intrinsically stretchable large-area organic field-effect transistors (OFETs). a,** Schematic illustration and key highlights of the fabricated complementary transistors. PSC, encap, and S/D represent organic semiconductors, encapsulation, and source/drain, respectively. **b,** Comparison of key performance metrics of intrinsically stretchable organic electronics. Three representative works of intrinsically stretchable organic electronics are highlighted as Gen-1[57], 2[34], and 3[48]. All references are listed in Supplementary Fig. 1. **c,** Schematic illustrations of the two critical interfaces with polymer semiconductors, the top encapsulation interface and the bottom low-k dielectric interfaces. The semiconductor layer can be swapped for different types of p-type and n-type stretchable polymer semiconductors (PSCs), including p-type DPPTT/BA and DPPTT/UC203M, and n-type BDOPV/BA. DPPTT, diketopyrrolopyrrole; BA, perfluorophenyl azide (PFPA) end-capped polybutadiene; BDOPV, tetrafluorinated benzodifurandione-based oligo(*p*-phenylene vinylene) with 2,2′-bithiophene. **d,** 2D schematic of the tri-layer channel interface. **d,** Optical microscope images of fabricated high-density transistors of 55,000 OFETs per cm² with channel lengths of 15 μm. The inset shows a zoomed-in optical microscope image of the high-density OFET array. **e,** Photo showing high-density soft OFET arrays forming conformal contact on a fingertip. **f,** Photo of large-area OFET arrays (4-inch wafer scale) under large deformation.

**OFET fabrication and transistor performance**

Previous stretchable OFET fabrication methods primarily relied on manual alignment and shadow masks to mitigate electronic performance degradation caused by photoresist patterning and etching. Those approaches, however, are not scalable as they introduce substantial device-to-device variations and are limited to large feature sizes[36,48,54]. Additionally, sensitive functional layers in the OFET were commonly transferred to fabricated devices to prevent degradation from solvent swelling, dissolution, or delamination caused by direct solution deposition. However, incomplete transfer and defects substantially limit throughput and device yield[48]. To overcome these challenges, we developed a PSC-compatible monolithic fabrication process using photolithography (**Fig. 2a**). A bottom-gate bottom-contact (BGBC) device configuration was used to minimize PSC exposure to solvents during fabrication. First, a stretchable conductive polymer, crosslinked poly(3,4-ethylenedioxythiophene) polystyrene sulfonate (PEDOT:PSS/P123), was crosslinked and patterned by oxygen plasma etching using a photoresist mask[58]. Then, a dual-layer dielectric with high-$k$ nitrile butadiene rubber (NBR, $k = 25$, ~320 nm thick) and a thin passivating low-$k$ styrene-butadiene-styrene (SBS, $k = 4$, ~50 nm thick) was deposited on top to achieve both low operational voltage and minimal hysteresis in the resulting transistor[48]. Previous literature[46,58] reported lengthy PMMA-assisted etching approaches for patterning the low-k dielectric material (**Supplementary Fig. 2a**). In this project, we adopted a direct photo-patterning method (**Supplementary Fig. 2b**) by leveraging thiol-ene chemistry with C=C bonds in SB[46]. Compared to the etching method, our direct photo patterning of low-k dielectric eliminated three fabrication steps and reduced processing time by 2 hours (**Supplementary Fig. 2**). Next, stretchable metallic carbon nanotube (CNT) source and drain electrodes were patterned with a bilayer poly(methyl methacrylate) (PMMA) and Cu-assisted lift-off process[58]. This high-resolution patterning technique for source/drain electrodes is critical for achieving high device density, enabling us to fabricate transistors with channel lengths as small as 2 μm (**Fig. 2b and Supplementary Fig. 3**), which is among the highest resolutions in stretchable OFET fabrication[34].

For the stretchable PSC channel layer, a polymer blend of poly-thieno[3,2-*b*]thiophene-diketopyrrolopyrrole (DPPTT) and perfluorophenyl azide (PFPA) end-capped polybutadiene (BA) was used due to its good mobility, robust solvent resistance from crosslinking, and good ambient stability[51]. A key innovation in our fabrication process is the self-aligned encapsulating (SAE) step of the PSC channel, which is also the final step in fabricating OFETs. The SAE is enabled by the superior adhesion between PSC (*i.e.*, DPPTT/BA) and SBS, where the direct photo-patterned thick SBS encapsulation also serves as an etching mask during plasma etching of PSC (**Fig. 2c, Supplementary Fig. 4**). The entire above fabrication process can be completed within a few hours in a university cleanroom (**Supplementary Fig. 5**). Combined with the maskless direct-write photopatterning technique, the SAE enables the rapid turnaround of various OFET designs and fabrication, thereby providing an effective platform for developing stretchable circuits for various applications.

Due to the processing improvements, the fabricated OFETs showed low operational voltages (3–5 V), high on/off ratios (>$10^5$), negligible hysteresis, and low off-state current (<1 nA) (**Fig. 2d, 2e**). A mobility of 0.235 ± 0.020 cm$^2$ V$^{-1}$ s$^{-1}$ was achieved (**Supplementary Fig. 6a, 6b**), on par with previously reported high-density stretchable OFETs[51]. Here, a bottom-gate, bottom-contact device structure was adopted to minimize exposure of the channel layer to solutions prior to encapsulation. Thus, good device uniformity and mobility were maintained at a channel length of 35 μm with negligible performance degradation during subsequent processing steps[51]. It should be noted that the effects of contact resistance become prominent when channel lengths decrease below ~15 μm (**Fig. 2f, Supplementary Fig. 3**), which can be further optimized using contact engineering (*e.g.*, band alignment between CNT electrodes and PSC channels[59,60]) to improve performance in small-channel OFETs.

We compared mobility before and after rinsing with toluene, the solvent used to dissolve SBS for the thick encapsulation layer, which is the only solvent that directly contacts and might degrade the channel layer. The mobility of OFETs showed a minimal drop of 4%. (**Supplementary Fig. 7a**). Additionally, the resulting OFETs demonstrated good operational stability in air, with the on-current decreased by less than 10% after more than one hour of continuous biasing at -5 V (**Fig. 2g**). Moreover, after being stored in ambient air for 2 weeks and nitrogen box for 9 months, mobility of fabricated transistors decreased by less than 21% (**Supplementary Fig. 7b**), indicating a good long-term stability.

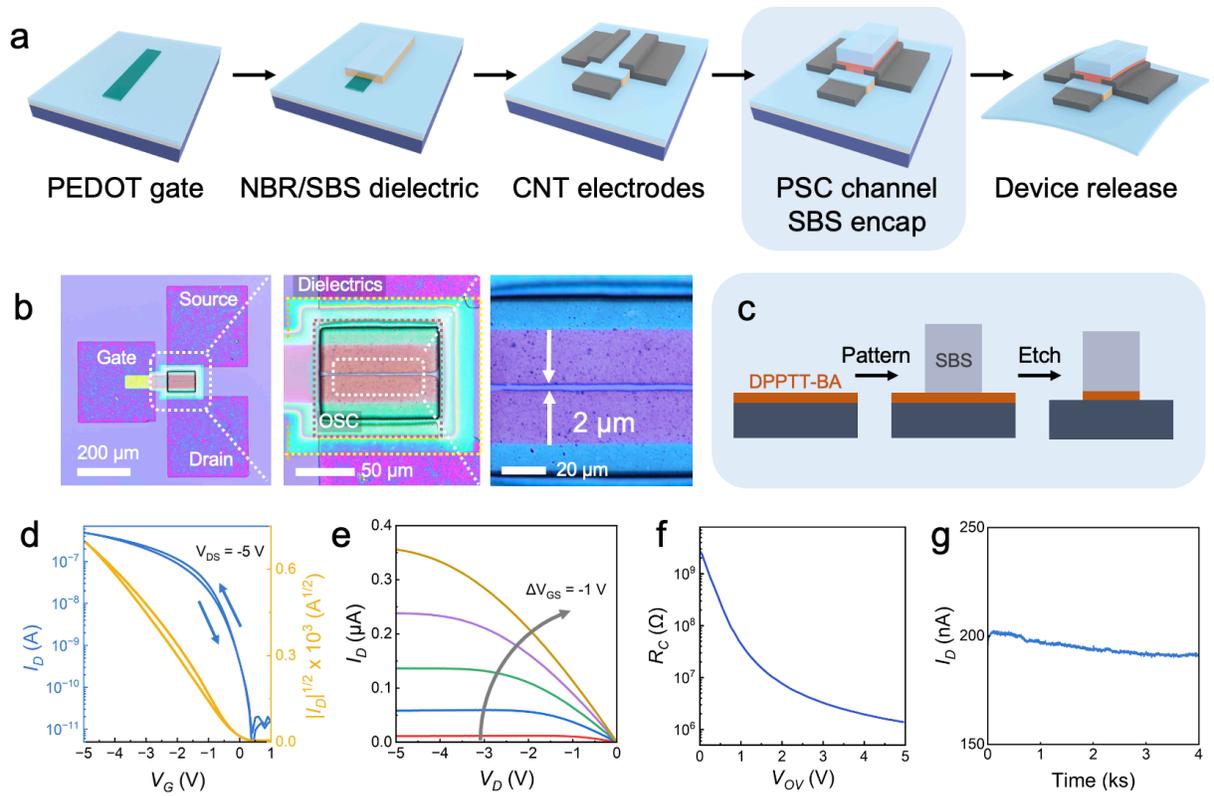

**Fig. 2. Photolithography-based fabrication and characterization of results stretchable OFETs. a,** Schematic of fabrication process of OFETs, with organic channel stage highlighted and emphasized in (c). PEDOT: poly(3,4-ethylenedioxythiophene); CNT: carbon nanotubes; NBR: nitrile rubber; SBS: styrene-butadiene-styrene; PSC: organic semiconductor. **b,** A representative fabricated transistor with high resolution channel (2μm) with different magnifications. **c,** A schematic highlighting the self-aligned encapsulation (SAE) process for patterning both the semiconductor and its encapsulation, shown in the step highlighted in (a). **d,e** Typical transfer (d) and output (e) characteristics of a fabricated OFET. $V_{GS}$ varies from 0 V (curve overlaps with the x-axis) to -5 V (highest $I_D$) in (e). W=380 μm, L=35 μm. **f,** Contact resistance of transistors under overdriving voltage ($V_{OV}$) from 0 to 5 V. $V_{OV} = |V_{GS}-V_{TH}|$. **g,** Real-time channel current ($I_D$) of an OFET biased under -5 V for both $V_{GS}$ and $V_{DS}$ over 1 hour under ambient conditions.

**Wafer-scale fabrication and high uniformity**

Developing processes with high uniformity across large areas and across different batches is crucial for the mass production of complex, large-area integrated circuits. Our monolithic high-resolution fabrication approach eliminates manual alignment and transfer-related processes, demonstrating the potential to address the above challenges in stretchable OFET fabrication. Our prepared large-area OFETs exhibit less than 20% on-current variation across a 4-inch diameter from the center (**Fig. 3a, 3b**). To assess batch-to-batch variation and device yield, we fabricated 2 cm × 2 cm chips in three separate batches, with 84 OFETs measured on each chip. All 252 measured OFETs were fully functional, corresponding to a yield rate close to unity. Moreover, the superimposed transfer curves for all 252 devices exhibit tight distribution and minimal variation (**Fig. 3c**). A detailed statistical analysis via histograms reveals a mean on/off ratio of >$10^4$ (**Fig. 3d**), an average mobility of 0.12 $cm^2$ $V^{-1}s^{-1}$ (**Fig. 3e**), a consistent negative threshold voltage with an average of -0.06 V (**Fig. 3f**), and a gain-width ratio of 0.37 nS/µm (**Fig. 3g**). The narrow distributions of all these parameters underscore the good reproducibility of our fabrication process, attributed to the monolithic, transfer-free and manual-alignment-free processes. Note that the mobility variations in OFETs made from different batches of DPPTT polymers were observed, which is attributed to the nature of the synthesis and purification processes of DPPTT.

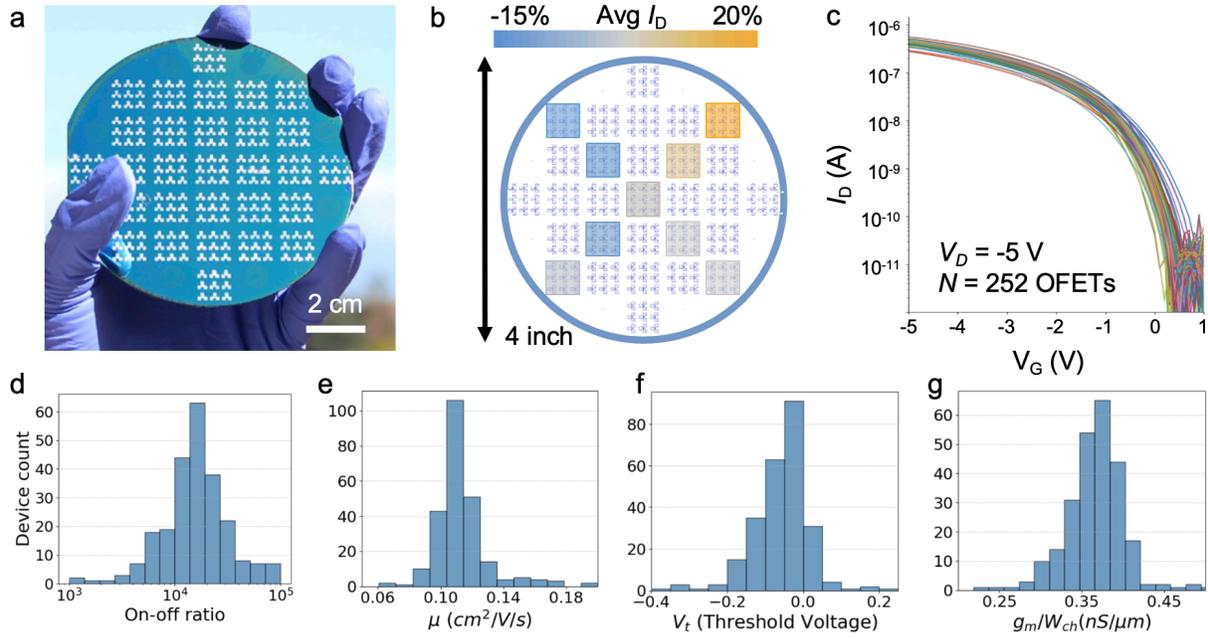

**Fig. 3. Fabrication uniformity of wafer-scale OFETs and batch-to-batch reproducibility. a,** Photo of monolithically fabricated transistors on a 4-inch wafer. **b,** Schematic illustration of current distribution across the 4-inch wafer. Representative 9 blocks of transistors diagonally distributed on the wafer were measured. Each square block containing nine transistors, and five representative OFETs were measured. The on-current was measured when $V_{GS}$ was swept from 1 to -5 V and $V_{DS}$ was biased at -5 V. The on-current of each block is rendered in colors that indicate its deviation from the average current across all blocks. **c,** Transfer characteristics of large-area stretchable 252 OFETs fabricated and measured across three batches of fabrication on 2 cm × 2 cm chips, 84 devices each. **d-g,** Histogram distribution of (d) on-off ratio, (e) mobility, (f) threshold voltage, and (g) normalized gain, of all 252 devices fabricated and measured.

**Stretchable p-type organic transistors and circuits**

Robust electronic performance under strain is important for biomedical and robotic applications. Built on entirely stretchable materials, our OFETs exhibit excellent stretchability with no observable cracks or delamination when stretched up to 100% strain (**Fig. 4a, 4b, Supplementary Fig. 8 a, b, c**). At 50% strain, our transistors experience a 33% drop in mobility along the strain direction. Compared to most reported transistors that have mobility drop by >50% under the same criteria, ours maintains good mobility on par with state-of-the-art (**Supplementary Fig. 8d**). In our case, the SBS encapsulation is ~3 μm thick, much thicker and stiffer than other functional layers of the devices (~500 nm in total). Therefore, it can function as an elastiff island that mitigates strain-induced performance degradation[36]. Building on this elastiff island effect, the actual channel length changes were 42% and -32% when the entire sample was stretched by 100% strain along and parallel and perpendicular to the channel directions, respectively. Upon release of the strain, the mobility recovered towards its initial value, confirming the elastic nature of the devices (**Fig. 4b**). Finally, the transistors were subjected to 1,000 stretching cycles to 25% strain. The mobility decrease is 3.2% and 0.1% respectively in two orthogonal directions, demonstrating the excellent mechanical robustness of our fabrication approach (**Fig. 4c**).

To demonstrate the capability of our process for more complex circuit integration, we fabricated pseudo-enhancement mode (pseudo-E) inverters, an important component in logic gates[61]. Eutectic gallium-indium (EGaIn) was used as the stretchable interconnect in the inverters due to its high conductivity (0.21 Ω/sq) and high stretchability of up to 100% strain[46]. The EGaIn electrodes were patterned using a similar photolithographic method as reported previously[58]. The inverter design utilizes a large channel width ratio difference—a factor of 125—between the driver and load transistors ($W_1/W_2$) to ensure a signal gain as high as 25 (**Fig. 4d**). Impressively, the inverter's voltage transfer curve remains stable even when the device is subjected to 100% tensile strain (**Fig. 4f**). To test the speed of our circuits, we fabricated 5-stage ring oscillators (ROs) by connecting five head-to-tail pseudo-E inverters (**Fig. 4e**) with another inverter not in the closed loop as the buffer to isolate circulating electrical signals from measurement setups. A typical sinusoid output generated by ROs is observed (**Fig. 4g**). We also observed the voltage-controlled oscillator (VCOs) behavior in our ROs. When $V_{DD}$ was changed from 3 V to 30 V, the output voltages of ROs increased from 0.6 V to ~30 V, and their frequencies increased from 320 Hz to 1.14 kHz (**Supplementary Fig. 9**). We note that this is the first time stretchable OFET RO circuits reached kHz frequencies. Considering the channel length is 15 μm, improving contact resistance at a small channel length should further substantially increase the operating frequency in the future. Ring oscillators with their component inverters having $W_1/W_2$ ratios of 20 and 10 were fabricated respectively, where $W_1$ and $W_2$ are defined in **Fig. 4d**. Both ROs showed a monotonic increase in frequency as the voltage increased (**Fig. 4h**). Moreover, we simulated our fabricated transistors and designed logic gates, including NAND and NOR (**Supplementary Fig. 10b,e**). The fabricated logic gates showed consistent performance with the simulated results. (**Supplementary Fig. 10c,f**).

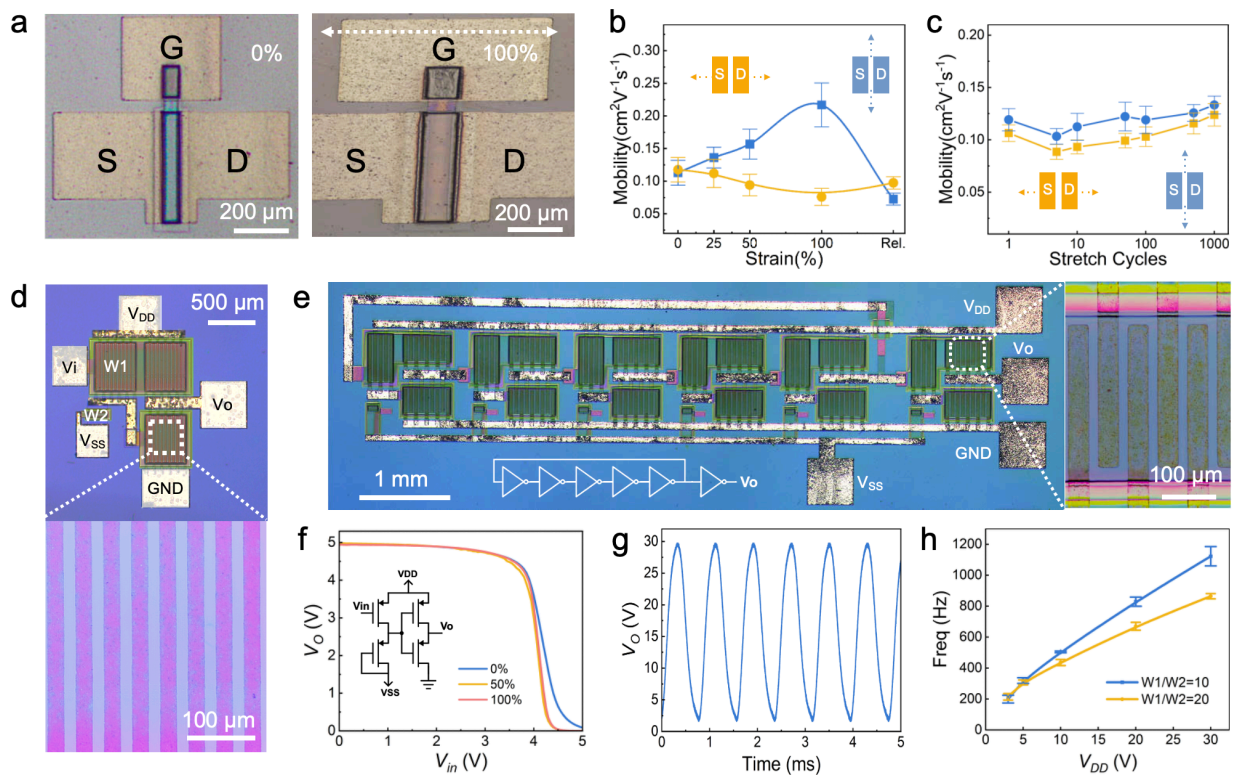

**Fig. 4. Stretchable p-type organic transistors and circuits. a,** Microscope images of a representative OFET at 0 and 100% strain. W/L=380/35 μm. The channel was under 42% strain when the entire substrate was stretched to 100%. Arrows indicate the stretch direction. G, S, and D represent gate, source, and drain electrodes, respectively. **b,** Mobility change under strain along (blue) and perpendicular (yellow) to channel width directions under strains of 0%, 25%, 50%, 100%, and 0% (released). Error bars show the standard deviation of $N = 12$ devices across three batches. **c,** Mobility change with stretching cycles up to 1,000 at strains of 25% along (blue) and perpendicular (yellow) to channel width directions. Error bars represent the standard deviations of $N = 18$ devices across three batches. **d,** Optical microscope images of a pseudo-E inverter with stretchable p-type OFETs; bottom: zoomed-in optical microscope image of the channel area. $W_1$ and $W_2$ represent the channel widths of the transistors shown in the text. The remaining two transistors share the same dimensions as transistor 1. Here, $W_1/L_1$=5000/10μm, $W_2/L_2$=200/35 μm. **e,** Optical microscope images of a 5-stage ring oscillator, with five repeating stages of inverters and one buffer at the output. The inset shows the circuit diagram. The microscope image on the right shows the zoomed-in image of the channel area, with $W_1/L_1$=2000/10μm, $W_2/L_2$=100/10μm. **f,** Transfer curves of a pseudo-E inverter under the strain of 0%, 50%, and 100%. The inset shows the schematic of a pseudo-E inverter shown in (d). **g,** Voltage output of a representative OFET ring oscillator shown in (e) with a frequency of 1.1 kHz. **h,** Frequency changes of ring oscillators under different $V_{DD}$ and $W_1/W_2$ ratios. Error bars are the standard deviation of $N = 12$ devices across three batches.

**Versatile fabrication process for other channel materials**

Building on the universal tri-layer channel interface structure (**Fig. 1c**), our OFET fabrication process is versatile for incorporating various PSC channel materials. In addition to the p-type DPPTT/BA materials, we applied an n-type stretchable PSC based on tetrafluorinated benzodifurandione-based oligo(*p*-phenylene vinylene) with 2,2′-bithiophene (F$_4$BDOPV-2T)[62]. The crosslinked F$_4$BDOPV-2T/BA film was used due to its high stretchability (of up to 100% strain) and peak mobility reported for stretchable electronics[63] of 0.28 cm$^2$ V$^{-1}$ s$^{-1}$. With our monolithic fabrication process, an optimized F$_4$BDOPV-2T/BA-based n-type OFET showed a maximum mobility of 0.184 cm$^2$ V$^{-1}$ s$^{-1}$, negligible hysteresis, and a high on/off current ratio of >10$^3$ (**Supplementary Fig. 6c, 6d, 11**).

Our current fabrication processes use an in-house-synthesized BA crosslinker. To increase the accessibility of the process to researchers without access to organic synthesis laboratories, we sought to replace BA with commercially available chemicals. We developed a polymer blend of a commercially available isoprene rubber-based polymer, UC-203M, with commercial DPPTT for stretchable OFET fabrication. The UC-203M forms an interconnected network with DPPTT after crosslinking (**Supplementary Fig. 12a**). Devices were prepared with ratios of DPPTT and UC-203M 3:7, 1:1 and 7:3. As the DPPTT weight ratios increase from 30% to 70%, more densely packed fiber structures were formed and higher on-currents at 10V supply voltage were observed (**Supplementary Fig. 12 b-g, Fig. 13 a, b**). With 7:3 DPPTT/UC-203M blend films, we achieved an average on-current of 2.89 μA on transistors with W/L ratio of 380/35 μm, similar to that of the DPPTT/BA devices fabricated in the same batch (3.06 μA) (**Supplementary Fig. 13c**). OFETs with DPPTT/UC-203M exhibit stretchability up to strain of 100% with mobility maintained at the same order of magnitude (**Supplementary Fig. 14**). These results demonstrate that stretchable OFETs fabricated from off-shelf chemicals, specifically DPPTT/UC-203M, can exhibit comparable electrical and mechanical properties to state-of-the-art systems, thereby opening up opportunities for rapid prototyping.

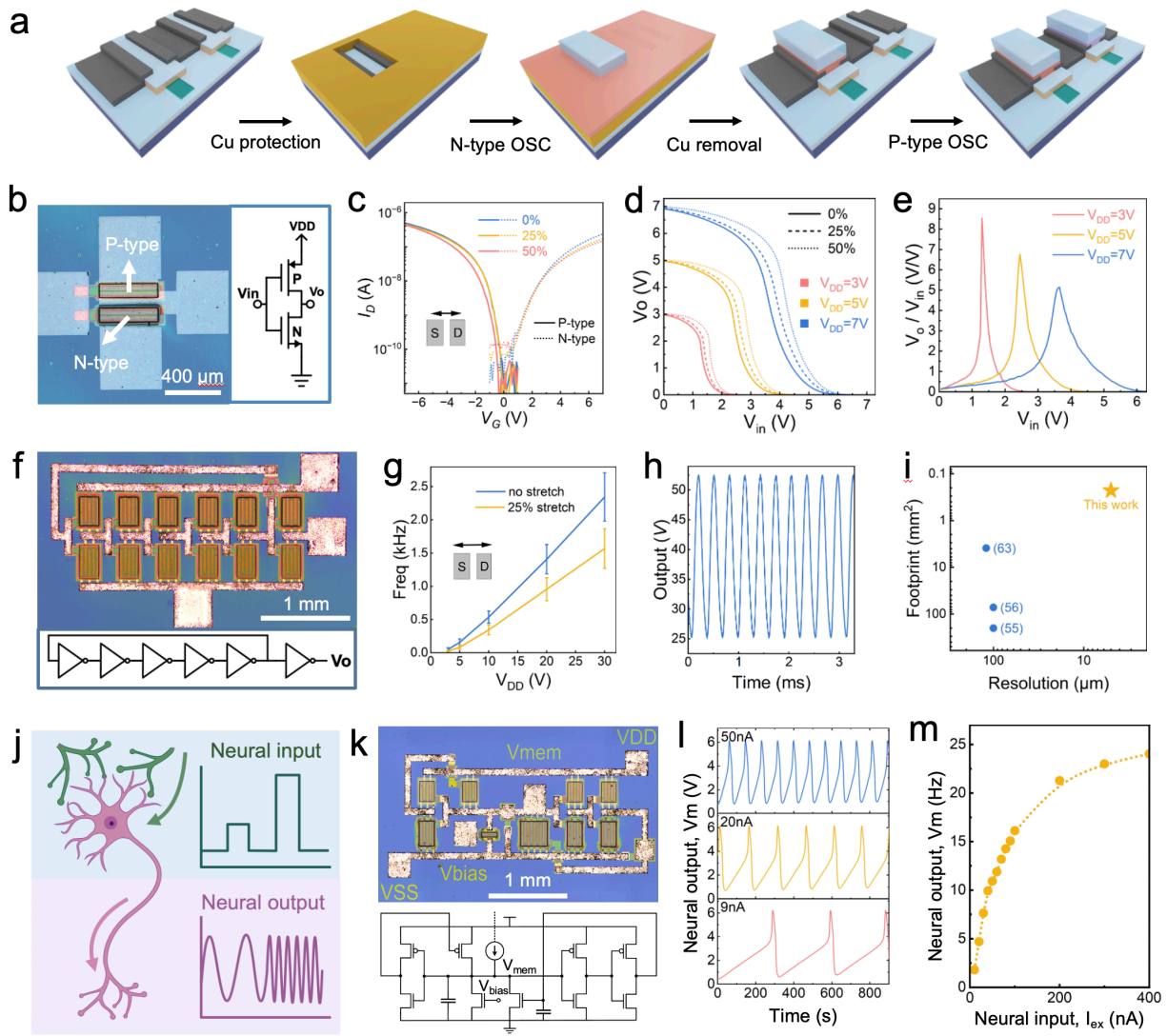

**Fig. 5. Stretchable organic complementary transistors and circuits. a,** Schematic illustration of the critical steps in the fabrication of complementary organic transistors and circuits, highlighting the development of copper isolation masks for patterning. **b,** Optical microscope image of a complementary organic inverter alongside its corresponding circuit schematic. **c,** Transfer characteristics ($I_D$−$V_G$) of representative p-type and n-type transistors on the same substrate, showing minimal change in performance under up to 50% applied strain. **d,** Transfer characteristics of the complementary inverter operating at different supply voltages ($V_{DD}$ = 3, 5, 7 V) and strains of the whole sample at 0, 25%, and 50%. **e.** Voltage gain of the inverters calculated from (d) at 0% strain. **f,** Optical microscope image of a fabricated 5-stage ring oscillator (RO) and its circuit schematic on the bottom. **g,** Oscillation frequencies of the RO as a function of $V_{DD}$ under the strain of 0 and 25%. Error bars represent the standard deviation of 4 devices. **h,** Output waveform showing an RO operating at 3.3 kHz with a 60 V supply voltage. **i,** Benchmark comparison of stretchable organic complementary circuits in the literature. The y-axis represents the footprint of active regions in the complementary inverting circuit, and the x-axis indicates the highest achieved resolution. **j,** An illustrative diagram of a neuron circuit with its representative input and output signals. **k,** A fabricated neuron circuit under the microscope and its schematic

design. **i,** Transient response of the membrane potential for different stimulation currents, from 9 nA to 50 nA. **m,** Relationship between output firing frequency and amplitude of injected current, showing the neuromorphic behavior.

**Stretchable organic complementary circuits**

Most intrinsically stretchable OFETs are constructed with only n- or p-type transistors. Complementary circuits are advantageous due to their lower static power consumption, increased functionality, and smaller circuit footprint, yet high-throughput fabrication processes for complementary stretchable circuits have not been developed. A critical challenge in organic complementary circuit fabrication is the degradation of the already-deposited n-type or p-type semiconductors during subsequent deposition and patterning steps. In this case, we found that our crosslinked SBS encapsulation layer can serve as an effective protection layer. Another common issue is the PSC residue on the dielectric surface from the patterning process, which can dope the subsequently patterned PSC and cause high off-current. Based on our transistor fabrication process, we developed an integrated process for sequential incorporation of n-type and p-type PSC into complementary circuits (**Fig. 5a**). To prevent the aforementioned residue problem, we developed a process that utilizes a sacrificial copper film as an isolation layer for the dielectric surface (**Fig. 5a, Supplementary Fig. 15**). With this method, a copper layer is first deposited and patterned to protect covered region from contamination and degradation from etching. Then, the n-type material is spin-coated and cross-linked on the entire sample, encapsulated by directly photopatterned thick SBS (~3-µm thick) and patterned by oxygen plasma etching. SBS will also experience thickness reduction due to etching; however, its functional part is largely conserved because it is significantly thicker than the n-type PSC (50 nm thick). Next, the copper mask is removed using $FeCl_3$ etchant. Lastly, the p-type semiconductor is deposited and patterned using the approach described in **Fig. 2a**. During the steps of p-type fabrication, the SBS layer fully protects the n-type PSCs from doping. The fabricated complementary transistors have symmetric geometry: apart from the channel material used, other layers have identical designs (**Supplementary Fig. 16**). This integrated process yields high-performance, high-density patterning of n- and p-type transistors on the same substrate, with transfer characteristics that are nearly identical to those of single-type devices fabricated on separate chips (**Supplementary Fig. 17**).

Lastly, we designed and fabricated some key building blocks of stretchable complementary logic circuits. First, we demonstrated the fabrication of a complementary inverter, the fundamental building block of logic gates, using p-type and n-type transistors that both exhibit stable transfer characteristics under tensile strain up to 50% (**Fig. 5b, 5c**). The inverter itself demonstrates excellent performance at a supply voltage as low as 3 V (**Fig. 5d**). Notably, transfer curves remain consistent when the circuits were stretched to 50% (**Fig. 5d**). The complementary inverter showed a high voltage gain of 11 at a $V_{DD}$ of 3 V and negligible hysteresis across all tested supply voltages (**Fig. 5e**, **Supplementary Fig. 18**). To evaluate the dynamic performance and integration density of our platform, we prepared multi-stage ring oscillators (ROs) (**Fig. 5f**). A key advantage of our complementary design over that of the pseudo-E is the 5-fold smaller device footprint (**Fig. 4e vs 5f**). This reduction in footprint leads to lower parasitic capacitance, enabling substantially faster circuit operation. Our 5-stage ROs demonstrate a record-breaking oscillation frequency for stretchable OTFT circuits, reaching over 3.3 kHz at a supply voltage of 60 V (**Fig. 5h**). This represents a nearly 67-fold improvement over the previous state-of-the-art OTFT circuits in terms of stage switching frequency[48]. These complementary ROs bring the performance of stretchable OTFT electronics into a regime for non-stretchable flexible circuits[64-67]. The circuits also exhibit impressive mechanical robustness and process scalability. The ROs maintain stable operation when stretched to 25%, with 35% decrease in frequency attributable to increased interconnect resistance. Crucially, our process achieves an excellent device yield of over 95%. This high yield, combined with the small device footprint compared to previous generations (**Fig. 5i**), underscores the potential of our platform for fabricating the high-density, complex circuits required for next-generation skin electronics. Our high-resolution, high-throughput complementary OFET fabrication process enabled circuits with complex functions that were previously challenging to achieve on 100-μm fabrication platforms. As a proof-of-concept, we designed and fabricated the first stretchable complementary OTFT neuron circuit (**Fig. 5j**), whose output is a pulse-like spike resembling a biological action potential and whose firing frequency is modulated by the input-controlled current. The firing frequency ranged from 3 Hz to 26 Hz as the input current varied from 9 nA to 500 nA (**Fig. 5k, 5m**). These neuromorphic circuits, built from entirely soft, stretchable materials, move beyond the signal conditioning capabilities exhibited by previous stretchable analog circuits (e.g. self-feedback inverter-based amplifiers), and display more sophisticated information abilities, that is temporal integration, thresholding, non-linear signal processing and event generation. Thanks to the event-driven nature of their operation, neuron circuits encode changes in the input signal, rather than its absolute value, so no spike is generated when the input signal remains constant. Transmitting only relevant events reduces data bandwidth, lowers power consumption, and avoids continuous sampling and digitization. All these characteristics make stretchable neurons promising platforms for future in-body computing and signal processing. Future studies can integrate these neuromorphic circuits with sensors to achieve sense-to-stimulate neural circuits.

**Conclusions**

In summary, we developed a widely applicable intrinsically stretchable organic transistor fabrication process adopting a tri-layer channel interface design. It is suitable for supporting a wide selection of materials and enables high-density, low-voltage operation with large-scale uniformity and a high yield rate. The resulting devices display a combination of high-performance characteristics, including negligible hysteresis, exceptional operational stability in air, and robust stretchability up to 100% strain. Our versatile fabrication process can be applied to various p- and n-type stretchable PSCs, including the commercialized UC-203M, which enables the use of all off-the-shelf materials. Furthermore, our method addresses major challenges in fabricating high-performance complementary circuits with both p- and n-type OFETs. We demonstrated inverters capable of operating at 3 V within a compact footprint, setting a new benchmark for area density in stretchable OTFT logic. The resulting ring oscillators achieved operating frequencies exceeding 3.3 kHz—a 67x improvement in stage switching speed over the previous state of the art. This leap in dynamic performance and integration density paves the way for the realization of complex, fully stretchable electronic systems. Building on this platform, we enabled the first stretchable OFET neurocircuit with pulse frequency adjustable by input-controlled current. The current limitation of transistors is their high contact resistance at small channel lengths (<15 μm). Future work will address this challenge, further broadening the utility of stretchable electronics.

**Methods**

**Materials.**

All materials were purchased from Sigma Aldrich unless otherwise mentioned. Solvents, including MEK (2-Butanone), isopropanol (IPA), toluene, cyclohexanone, and cyclohexane, were all used as received. Aqueous PEDOT:PSS dispersion CLEVIOSPH 1000 was purchased from Ossila and used as received. Poly(Styrene-Butadiene-Styrene D1102) (SBS) was purchased from Kraton Polymers, Poly(styrene-ethyle-butylstyrene H1211) with the volume fraction of poly(ethylene–co-butylene) of 82% (SEBS) was provided by Asahi Kasei. Dowsil SYLGARD 184 Silicone Elastomer (PDMS) was purchased from DOW (LOT # H047NBD001). PMMA 495k A4 and MF-319 photoresist developer were purchased from Kayaku Advanced Materials. Photoresist AZ 1512 was purchased from Microchemicals. MG Chemicals 415-1 L Ferric chloride copper etchant solution was purchased from MG Chemicals. Poly(acrylonitrile-co-butadiene) with acrylonitrile 37–39 wt% (NBR), dextran with Mr ~100,000, PMMA 495k A4, Gallium–Indium eutectic (EGaIn), Pentaerythritol tetrakis(3-mercaptopropionate) (PETMP), phenylbis(2,4,6-trimethylbenzoyl)phosphine oxide (BAPO), lithium phenyl-2,4,6-trimethylbenzoylphosphinate (LAP) and Poly(ethylene glycol)-block-poly(propylene glycol)-block-poly(ethylene glycol) diacrylate with average Mn ~5800 (P123 DA) were purchased from Sigma-Aldrich. P3-SWNTs (CNT) was purchased from Carbon Solutions. CNT dispersion solution was prepared as reported previously[58]. UC-203M was purchased from Kuraray Liquid Rubber. Commercial DPP-DTT, PDPP2T-TT-OD (DPPTT) can be purchased from Ossila. DPPTT with high performance (mobility ~0.26) was synthesized following a previously reported recipe[6,51,57]. F$_4$BDOPV-2T was synthesized according to previous reports[62]. For DPP-DTT, Mn = 54.5 k and $M_w$ = 138 kg mol$^{-1}$; for BDOPV, $M_w$ = 118.0 kg mol$^{-1}$.

**Fabrication process of intrinsically stretchable transistors.**

1. **Sacrificial layer:** Silicon wafers with a 300 nm SiO$_2$ layer were used as the initial substrates. After an oxygen plasma clean (150 W, 100 s), an aqueous dextran solution (70 mg/mL) was spin-coated at 1000 rpm for 60 s and subsequently baked at 150 °C for 30 min to form the sacrificial layer. Finally, the prepared wafers were diced into individual chips for device processing.

2. **Substrate:** The stretchable substrate was fabricated by spin-coating a mixture of SBS (80 mg/mL dissolved in toluene), PETMP crosslinker of 2% weight ratio to that of SBS (2 wt% w.r.t. SBS), and BAPO photoinitiator (1 wt% w.r.t. SBS) onto prepared chips. The spin-coating recipe consisted of a two-step process: 800 rpm for 70 s, followed by 3000 rpm for 20 s. The resulting film was then photo-crosslinked inside a nitrogen-filled glovebox via UV exposure (Electro-Lite Corporation ELC-500, 30mW/cm² at 365nm) for 5 minutes, followed by a post-exposure bake at 150 °C for 20 minutes. Finally, the un-crosslinked polymer was removed by rinsing with cyclohexane, and the substrate was baked at 120 °C for 20 minutes to remove residual solvent.

3. **Gate:** A PEDOT:PSS solution with 30 mg/mL P123 DA and LAP (3 wt% w.r.t. P123 DA) was spin-coated onto the substrate at 3500 rpm for 70 s. The film was immediately UV-crosslinked (ELC-500, 30mW/cm² at 365nm) for 6 minutes inside a glovebox. Then it was treated with nitric acid (6 M) for 10 seconds in air, rinsed with deionized water, and baked at 130 °C for 10 minutes. This conductive layer was then patterned using standard photolithography: AZ 1512 photoresist was spin-coated (3000 rpm, 45 s), exposed to a 90 mJ/cm² UV dose, and developed using MF-319. All lithography in this project were done with Durham Magneto Optics ML3 MicroWriter direct write machines. The unprotected PEDOT:PSS was subsequently removed with an oxygen plasma etch (100 W, 150 s). Finally, the photoresist mask was stripped by sonicating the sample in acetone for 12 seconds, followed by a final bake at 130 °C for 5 minutes.

4. **Dielectric:** A solution of NBR (40 mg/mL in cyclohexanone) containing PETMP (4 wt% w.r.t. NBR) and BAPO (4 wt% w.r.t. NBR) was spin-coated at 700 rpm for 70 s, followed by 3000 rpm for 20 s. The film was patterned via photolithography with an approximate UV dose of 150 mJ/cm². (It should be noted that the optimal dose varied between 90–400 mJ/cm² depending on the specific NBR batch, necessitating preliminary dose tests). After a post-exposure bake at 65 °C for 2 minutes, the pattern was developed by spin-rinsing with methyl ethyl ketone (MEK). A final hard bake at >110 °C for 3 minutes completed the high-k dielectric layer. Next, a low-k material, a solution of SBS (10 mg/mL in toluene) with PETMP (4 wt% w.r.t. SBS) and BAPO (4 wt% w.r.t. SBS) was spin-coated at 1500 rpm for 60 s, then 4000 rpm for 20 s. This film was exposed to a 500 mJ/cm² UV dose and subsequently baked at 65 °C for 2 minutes. The layer was then developed by spin-rinsing with cyclohexane at 1500 rpm and finalized with a bake at 110 °C for at least 2 minutes.

5. **Source and Drain Electrodes:** A PMMA sacrificial layer was spin-coated at 700 rpm for 70 s, then 3000 rpm for 20 s, and dried in a desiccator for 1.5 hours. Baking at elevated temperature was not adopted to avoid thermal crosslinking between PMMA and the layers below. Then, a 150 nm copper hard mask is evaporated at 1.2 Å/s. The copper was patterned using conventional photolithography (AZ 1512 resist, 95 mJ/cm² exposure) followed by a 10-second wet etch in 1:20 diluted iron(III) chloride in deionized water. After flood exposing the photoresist (American Ultraviolet Cool Cure Chamber) for 4 seconds and stripping with MF-319, the underlying PMMA was briefly etched in acetone (5 s) to create an undercut for cleaner lift-off. CNT was then spray-coated onto the heated substrate (80 °C) until sheet resistance between adjacent test points measured less than 2 kΩ, followed by E-Beam evaporated contact layer Pd/Au (4 nm / 3 nm) at 0.5 Å/s. The final electrode pattern was achieved by a 2-hour soaking in acetone to liftoff PMMA/Cu and bake at 100 °C for 5 minutes to remove solvent.

6. **(Steps for complementary transistors) N-type channel:** A 160 nm copper isolation film was e-beam evaporated at 1.2 Å/s, and patterned and wet-etched as described previously in step 5, such that copper did not cover areas designated for n-channels. The substrate was then baked in a glovebox at 100 °C for 3 minutes, with the temperature ramped gradually to prevent cracking of the copper film. For the semiconductor ink, BDOPV was dissolved in chlorobenzene (5 mg/mL) by stirring at 85 °C for over 4 hours. A separate solution of BA crosslinker (25 mg/mL in chlorobenzene) was prepared and filtered through 1-μm PTFE filters before use. The two solutions were blended (5:1 BDOPV:BA solutions, v/v) and spin-coated onto the substrate at 1000 rpm for 60 s. Following deposition, the n-type semiconductor film was crosslinked with 254 nm UV light (using E-Series UV-A Blacklight Lamp from Spectro-UV) for 10 minutes and annealed at 85 °C for 1 hour. The n-type channel was then encapsulated with a photopatternable SBS dielectric. A toluene solution of SBS (150 mg/mL), PETMP (4 wt% w.r.t. SBS), and BAPO (2 wt% w.r.t. SBS) was spin-coated at 3000 rpm for 45 s, exposed to a 150 mJ/cm² UV dose using MicroWriter ML3 in air, baked at 70 °C for 2 minutes, and developed with cyclohexane at 1500 rpm. Uncovered BDOPV was removed by subsequent oxygen plasma etch at 100 W for 70 s.

Finally, the copper mask was stripped by immersing the sample in an iron(III) chloride solution.

7. **P-type channel:** The substrate was baked in a glovebox at 110 °C for 3 minutes to remove solvents. DPPTT was dissolved in chlorobenzene (5 mg/mL) by stirring at 85 °C for over 4 hours. A separate solution of BA crosslinker (25 mg/mL in chlorobenzene) was prepared and filtered (1 μm PTFE) before using. The two solutions were blended (5:1 DPPTT:BA solutions, v/v) and spin-coated onto the substrate at 1000 rpm for 60 s in a glovebox. Following deposition, the p-type semiconductor film was crosslinked and annealed by baking at 140 °C for 1 hour in a glovebox, with temperature gradually ramped up over 2 minutes and slowly ramped down over 20 minutes. A toluene solution of SBS (150 mg/mL), PETMP (4 wt% w.r.t. SBS), and BAPO (2 wt% w.r.t. SBS) was spin-coated at 3000 rpm for 45 s, exposed to a 150 mJ/cm² UV dose with MicroWriter ML3, baked at 70 °C for 2 minutes, and developed with cyclohexane at 1500 rpm. The exposed DPPTT was removed by subsequent oxygen plasma etch at 100 W for 42 s with March Instruments PX-250 Plasma Asher.

8. **(Steps for large circuits) Interconnects:** The EGaIn electrodes were fabricated using a single lift-off process. First, photoresist A1512 was spin-coated at 3000 rpm for 45 s, exposed to a 100 mJ/cm² UV dose using MicroWriter ML3, post-baked at 65 °C for 1 min, and developed in MF-319. A Titanium/Gold (5 nm/35 nm) adhesion layer was then deposited via e-beam evaporation at 0.5 Å/s, respectively. To perform stencil printing, a piece of PDMS was wetted with EGaIn and brushed over the samples with a photoresist liftoff mask. Finally, EGaIn was stencil printed on the sample and then slowly lifted off in acetone for over 4 hours without agitation. Remaining EGaIn was rinsed off with Acetone after 4 hours.

**Device characterization**

All electrical measurements were performed in ambient air. Microscope images were taken with Nikon LV100ND. The capacitance of the dielectric film was measured by a Keysight E4980A Precision LCR Meter. The current-voltage characteristics of transistors and transfer curves of inverters were recorded using a Keithley 4200 Parameter Analyzer. For the p-type-only inverter circuits, a DC power supply was used to provide the pull-up voltage. The dynamic performance and oscillation frequency of the ring oscillators were characterized using a digital oscilloscope.

**Circuit simulation**

Computer simulations were done with Cadence Virtuoso. Fabricated transistors were first fit into a model for thin film CNT parameters defined by measurement: channel width W, channel length L, dielectric capacitance $C_{ox}$, S/D overlap $L_{OV}$, threshold $V_{th}$ (by taking the average value calculated from histogram in **Fig. 4**); and variable parameters: contact resistance $R_C$, subthreshold swing SS, mobility μ, channel length modulation λ and gate dependent mobility $\gamma$. Then circuit simulations were performed with transistor models with the defined model.

Circuit simulations were performed in the Cadence Virtuoso environment. First, a device model was created by fitting experimental data from our fabricated transistors to an existing thin-film transistors[68]. The model used two sets of parameters:

- **Fixed Parameters.** These were defined by direct measurement and included channel width (W), channel length (L), gate dielectric capacitance ($C_{ox}$), source/drain overlap length ($L_{OV}$), and the average threshold voltage ($V_{th}$) extracted from the statistical data in **Fig. 4**.
- **Fitting Parameters.** These variables were adjusted to calibrate the model to the measured device characteristics. They included contact resistance ($R_C$), subthreshold swing (SS), charge carrier mobility (μ), channel length modulation (λ), and the gate-voltage-dependent mobility coefficient (γ).

Once calibrated, these transistor models were used to simulate the performance of the complete inverter, ring oscillator and logical circuits.

**Material characterization**

Nanomechanical images of the polymer thin films were generated via the Bruker Dimension Icon AFM with a Nanoscope V controller using SCANASYST-AIR cantilevers from Bruker AFM Probes. The cantilever force constant was calibrated by thermal tuning to $k = 0.43$ N/m[69]. The nominal tip radius of 5 nm was used. The experiments were performed at a setpoint of 500 pN, a peak force frequency of 2 kHz, and an amplitude of 30 nm. The scans were performed at a resolution of 256 by 256 with a scan rate of 0.9 Hz per line. The measured nanomechanical imaging data and histograms were processed using *Gwyddion* SPM software.

**Data availability**

Data that support the findings of this study are available from the corresponding author upon reasonable request.

**Acknowledgment**

Y.Yuan, Y. Yao, and Z.Y. acknowledge funding and support from the Shoucheng Zhang Fellowship. Y.Yuan acknowledges support from Lu Zhang. C.Z. acknowledges funding from an F32 fellowship from the National Institute of Biomedical Imaging and Bioengineering of the National Institutes of Health (F32EB034156). Z.B. is a Chan Zuckerberg Biohub San Francisco investigator and an Arc Institute innovation investigator. Z.B. acknowledges support from the Tianqiao and Chrissy Chen Ideation and Prototyping Lab and Stanford Wearable Electronics Initiative (eWEAR) seed funding. Authors thank Ena Luis for helpful suggestions to writing the paper. Part of this work was performed at the Stanford Nano Shared Facilities (SNSF), supported by the NSF award ECCS-2026822.

**Author contributions**

Y. Yuan, C.Z., M.R. and Z.B. designed the project and experiments. Y. Yuan, C.Z. and M.R. carried out experiments and collected related data. Y. Yuan, C.Z., M.R., Y.N., D.Z., C.W., H.K., K.C., Z.Y., W.W., Y. Yao, and J.Z. fabricated OFETs and related circuits. Z.S., R.M., Y.S., J.S., Q.L., H.W., and J.P. provided the necessary materials for the project. Y.Yuan, C.Z., and Z.H. performed electrical characterizations. L.F.M. and R.W. performed AFM characterization for the surface morphology of semiconductor materials. Y. Yuan, C.Z., and Z.B. wrote the manuscript. All authors reviewed and commented on the manuscript.

**Competing interests**

The authors declared no competing interests.

**Additional Information**
All data are available in the manuscript or Supplementary Information.